\documentclass[amsmath,amssymb,aps]{revtex4}

\usepackage{graphicx}
\usepackage{dcolumn}
\usepackage{bm}

\begin{document}


\title{Power Law Multi-Scaling of Material Strength}

\author{Alexander M. Korsunsky}
\email{alexander.korsunsky@eng.ox.ac.uk}
\affiliation{Department of Engineering Science, University of Oxford\\ 
Parks Road, Oxford OX1 3PJ, UK.}

\date{\today}

\begin{abstract}

Power law is one of the the simplest forms of the relationship between different variables of a system. It leads naturally to the introduction of compound parameters describing physical properties of the system. Often one of the variables of interest is the object dimension, or time. The prevalence of a simple power law over the entire range of dimensions or times can be helpfully interpreted as size or time independence of the corresponding compound physical parameter of the system. However, it is also often found that a simple power law only persists for some extreme values, e.g. for very large and/or small sizes, or very short or long times. Transitions between regimes of different power law asymptotic behaviour are encountered frequently in the description of a wide variety of physical systems. While asymptotic power law behaviour may often be readily predicted, e.g. on dimensional grounds, the evaluation of the relationship between system parameters in the transition range usually requires laborious detailed solution. To obviate this difficulty we introduce, on rather general basis, something we refer to as the merging, or 'knee' function. The function possesses sufficient flexibility to describe transitions of various sharpness. To demonstrate its usefuness, the merging function is applied to a variety of well-known scaling laws in the mechanics and strength of materials and structures.  

\end{abstract}

\maketitle

\section{\label{sec:intro}Power law multi-scaling in deformation and fracture}

Power laws play an important role in the study of size dependence of various physical properties. Barenblatt \cite{baren} and Bazant \cite{bazant} use functional analysis to demonstrate that, in the absence of an inherent length scale, the dependence of an arbitrary physical parameter of the system on its size must obey a power law. If two physical quantities are related by a power law, $y = C x^\alpha$, then the parameter $C=y x^{-\alpha}$ can be thought of as invariant in $x$. If $x$ represents system size, then $C$ is said to be size independent, or scale invariant. This parameter is often thought of as a property of the system, and it dimension is compounded from those of the quantities $x$ and $y$. 

However, in practice it is the deviation from the simple power law descriptions that is of particular concern. The terms 'size effect' or 'scale dependence' are conventionally used precisely when a particular power law no longer applies. That is, the parameter $C$, the system property, is no longer constant, and has to be thought of as a function of $x$.

In the present context we focus the attention on the size dependence of material strength. The precise nature of the physical parameter $y$ representing 'strength' depends on the context: it could be a critical value of stress, or stress rate, stress intensity factor, hardness, etc. We consider several instances of power multi-scaling, i.e. the transition between distinct power laws that persist over certain limited ranges of size. We then pose and answer the key question about efficient description of this type of multi-scaling.

By way of illustration, consider a plane strain crack of length $2a$ in an infinitely extended elastic solid subjected to a nominal remotely applied stress $\sigma$ the derivation of the Griffith \cite{grif} propagation criterion begins with the following expression for the total system free energy
\begin{equation}
F(a,\sigma)=4\gamma - \frac{\pi(1-\nu^2)}{E}\sigma^2 a^2.
\end{equation}
It is of particular interest to note the presence of two terms in the above expression displaying the dependence on different powers of the crack dimension, $a$. The first of the two terms represents the additional surface energy associated with the freshly created crack surfaces, proportional to the surface energy density, $\gamma$. The second term is negative and represents the reduction in the strain energy stored in the system due to the load relieving effect of the crack.

The presence of the two energy terms with different size scaling behaviour ensures that the free energy of the system is not a monotonic function of the size. The energy rather displays an extremum, found by setting the derivative with respect to size to zero,
\begin{equation}
\frac{dF}{da}=4\gamma - \frac{2\pi(1-\nu^2)}{E}\sigma^2 a=0.
\end{equation}
Hence, the critical combination of crack size and applied stress for crack propagation is found to be
\begin{equation}
\sigma^2\pi a=\frac{2\gamma E}{1-\nu^2}=K_{Ic}.
\end{equation}
where $K_{Ic}$ is considered to be a material property. This compound material parameter has unusual dimension of ${\rm MPa}\sqrt{\rm m}$, and was found to be constant for defects of sufficiently large size. 

According to the linear elastic fracture mechanics (LEFM) criterion, the scaling law for the failure stress as a function of crack size is given by
\begin{equation}
\sigma = \left(\frac{K_{Ic}}{\pi a} \right)^{\frac12}.
\end{equation}
Consequently, for specimens with cracks of length $a$ and $a_1$ the ratio of strengths can be written as
\begin{equation}
\frac{\sigma}{\sigma_1}=\left( \frac{a}{a_1}\right)^{-\frac12}.
\end{equation}
This is the most widely known scaling law in linear elastic fracture mechanics: the failure stress decreases as the inverse square root of the crack dimension. This power law behaviour is represented on the bilogarithmic scale of failure stress vs specimen size by a straight line with gradient $-1/2$.

It is immediately apparent that a paradox arises when cracks of progressively smaller size are considered. According to the above LEFM scaling law, the specimen strength would continue to increase without limit, as smaller and smaller crack sizes are considered. This is clearly in contradiction with the observations, since even most carefully prepared and defect-free specimens of any material always possess finite strength. For very small defect sizes, and particularly when no discernible flaw can be detected, a stress-based failure criterion must be used instead:
\begin{equation}
\sigma=\sigma_0,
\end{equation}
or, in the explicit form of a power law relationship between $\sigma$ and $a$,
\begin{equation}
\sigma=\sigma_0 a^0.
\end{equation}

The distinction between the two extremes of behaviour within the framework of LEFM is sometimes referred to as the long crack and the short crack regimes, respectively. In the long crack regime the inverse square root scaling with crack size is expected. In the short crack regime the specimens must possess a limiting stress value associated with alternative failure phenomena, e.g. yield. In order to capture the strength scaling behaviour over the entire range of crack length, one requires a combination of the two criteria, or a {\it multi-scaling} description.

Some simple but fundamental questions arise. 

Firstly, given that each power law applies in its respective extreme of flaw size, namely, the stress-based failure criterion when $a<<a^*$, and the fracture mechanics based criterion when $a>>a^*$, how can the transition between the two power law regimes be described?

Secondly, in formulating the above statement one is forced to introduce the parameter $a^*$ in order to express the asymptotic behaviour. What is the nature and the value of parameter $a^*$? 

These questions are typical of many situations in science and engineering, and are usually given {\it ad hoc} answers that depend on the nature of the system being studied. In particular, the usual procedure is to develop a more refined and detailed model that is capable of describing the 'interaction' between the two asymptotic regimes, but invariably requires significantly greater effort to solve. Requirements naturally applied to such a model would certainly include the conditions that in the respective limiting regimes the model should converge to the simple power law descriptions. 

The approach proposed in this paper aimis to obviate as far as possible the need for developing detailed solutions for the transition region. Rather, we ask whether it is possible to make some simple assumptions about the nature of the transition and to derive a general form of the 'merging function' between the two regimes. When once such form is established, the merging function can be calibrated using limited amount of experimental data, and will provide a very simple and portable expression of the multi-scaling power law response of the system.

In the following section we develop a form of the merging function on the basis of some fairly arbitrary assumptions. We do not contend at this point to claim that this is the correct, the best or the only form of such function that can be identified. Instead we seek to obtain a formulation that can be applied to a variety of situations in order to check its capability of describing the multi-scaling power law transition. We then focus our attention on illustrating the applicability of the proposed function in a wide variety of situations drawn from the context of deformation and strength of materials.

\section{\label{sec:deriv}'Knee' function description of the transition between two different power law regimes}

Consider a physical system described by a relationship between two parameters, $x$ and $y$, that dispays the following asymptotic behaviour:
\begin{equation}
\left\{
\begin{array}{c}
 x << x_0, \quad (y/y_0)=(x/x_0)^{\alpha}, \\
 x >> x_0, \quad (y/y_0)=(x/x_0)^{\beta}.
\end{array}
\right.
\end{equation}
The exact nature of the transition is not known. However, to make further progress we shall assume that there exists a non-dimensional parameter of the system, in the form of a product of powers of $y/y_0$ and $x/x_0$, which is given simply by a sum of two power law functions of $x/x_0$:
\begin{equation}
2(y/y_0)^\gamma=(x/x_0)^{\gamma\alpha}+(x/x_0)^{\gamma\beta}.
\end{equation}
Expressing the original parameter, $y$, from this form, we obtain:
\begin{equation}
(y/y_0)=(x/x_0)^\alpha \left[\frac{(1+(x/x_0)^{\gamma(\beta-\alpha)})}{2}\right]^{1/\gamma}.
\end{equation}
The division by 2 within the brackets ensures that the transition between the two regimes occurs at the point $x/x_0=1$, $y/y_0=1$.

We further consider a special case when one of the power law regimes corresponds to a constant, i.e. $\alpha=0$. In this case it is convenient to describe the scaling behaviour by the function:
\begin{equation}
(y/y_0)=\left[(1+(x/x_0)^{\gamma\beta})\right]^{1/\gamma}.  
\end{equation}
One particular situation often encountered in practice is when both $\beta$ and $\gamma$ are negative, $\beta=-b$, $\gamma=-c$, leading to
\begin{equation}
(y/y_0)=\frac{1}{\left[(1+(x/x_0)^{bc})\right]^{1/c}}.
\label{eq:knee}
\end{equation}
The above equation often describes the intraction between two different physical mechanisms that results in an {\it increase} of a particular property $y$ (strength, say) over the base value, $y_1$. It is then appropriate to write th above expression in the form
\begin{equation}
(y-y_1)/y_0=\frac{1}{\left[(1+(x/x_0)^{bc})\right]^{1/c}}.  
\end{equation}
leading to 
\begin{equation}
y=y_1+\frac{y_0}{\left[(1+(x/x_0)^{bc})\right]^{1/c}}.  
\end{equation}

\begin{figure}
\centerline{ \includegraphics[height=8.cm]{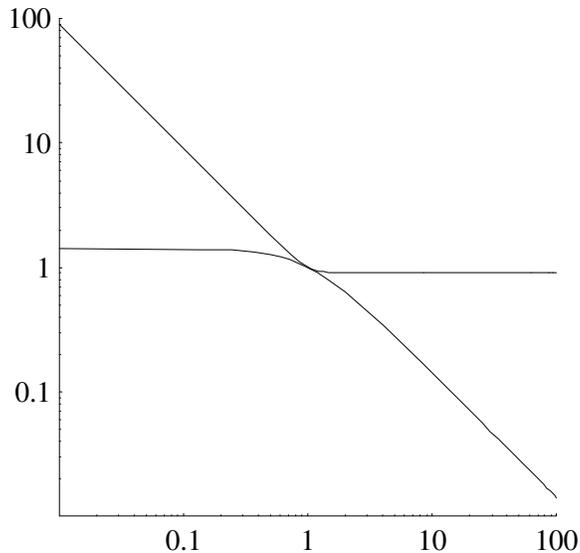} }
\caption{Illustration of multi-scale power law transitions described by the 'knee' function with different sharpness parameters.}
\label{fig:one}
\end{figure}

The formulae derived in this section are rather simple, and contain a relatively small number of parameters. However, they are found to describe successfully a wide range of multi-scale power law relationships in the science of material strength. In the following sections we provide illustrations of this fact, and then discuss implications and draw conclusions at the end. 


\section{\label{sec:dugdale}The transition between stress and toughness-controlled strength}

The existence of the two regimes of strength scaling described in the introductory section, namely, the toughness and stress controlled strength, already became apparent in the early experiments of Irwin \cite{irwin}. He conducted experiments on thin (0.8mm) large sheet of aluminium 7075 alloy containing central slits normal to the tensile loading direction. Irwin calculated the net section stress at instability, accounting for finite width of the sheet, and plotted this stress as a function of the crack (slit) length $2a$. Irwin's results clearly demonstrated the inverse square root behaviour predicted by LEFM, as illustrated by the dropping portion of the curve shown in Fig.2. However, it is also apparent from Irwin's results that for shorter cracks there exists a limiting value of stress that is independent of the crack length $2a$, provided it does not exceed some threshold value, $2a<2a_1$. The two failure criteria with different scaling behaviour need to be combined to obtain a description that is valid over the entire range of crack lengths considered.

Fig.3 contains the Irwin data re-plotted in the bi-logarithmic scale to highlight the power law multi-scaling behaviour. The data markers appear alongside the curve that represents the merging function of the type introduced in the previous section. The form of the function is chosen to satisfy the asymptotic behaviour for exteme values of the crack length parameter $a$, and is given by  
\begin{equation}
\frac{\sigma}{\sigma_1}=\frac{1}{(1+(2a/2a_1)^n)^{1/(2n)}}.
\label{eq:irwinfun}
\end{equation}
The best values of the other parameters appearing in this expression are given in Fig.2 and Fig.3.

The two-criterion failure condition introduced here occupies an important place in fracture mechanics analysis. It arises naturally within the elasto-plastic fracture process zone model, that is alternatively known as cohesive zone, bridging zone, or strip yield model, and is common;y associated with the names of Barenblatt \cite{bar} and Dugdale \cite{dug}. This model is an important tool in the analysis of size effects in fracture.

\begin{figure}
\centerline{ \includegraphics[height=8.cm]{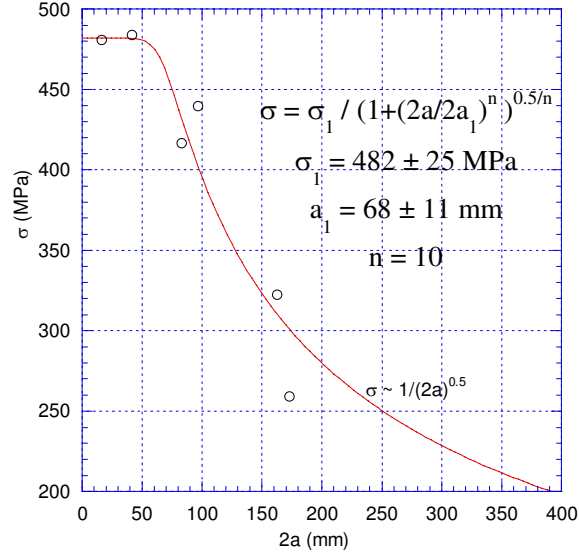} }
\caption{The Irwin et al. \cite{irwin} data for the net failure stress as a function of the total slit length $2a$, plotted in linear coordinates, together with the merging function describing the transition.}
\label{fig:two}
\end{figure}

\begin{figure}
\centerline{ \includegraphics[height=8.cm]{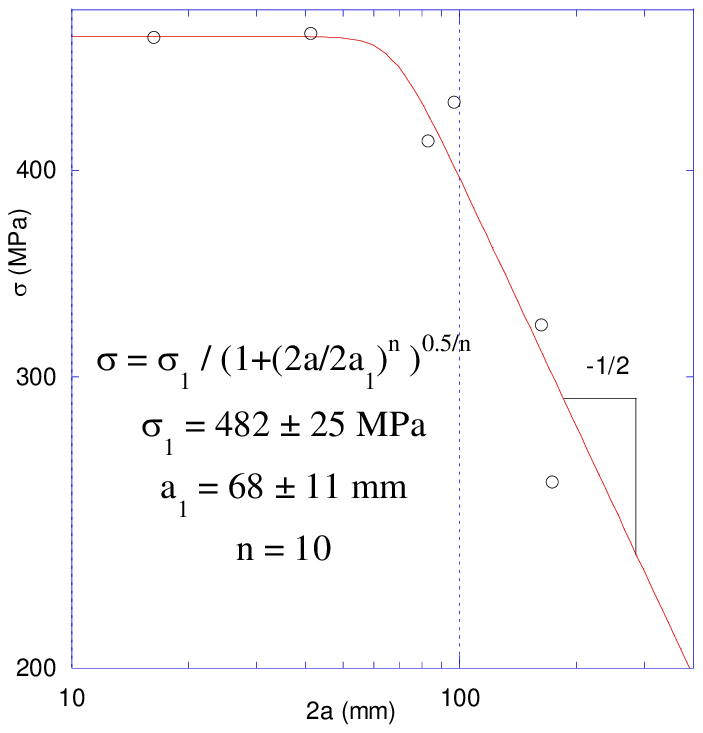} }
\caption{The Irwin et al. \cite{irwin} data for the net failure stress as a function of the total slit length $2a$, plotted
in bi-logarithmic coordinates, together with the merging function describing the transition.}
\label{fig:three}
\end{figure}

\begin{figure}
\centerline{ \includegraphics[height=8.cm]{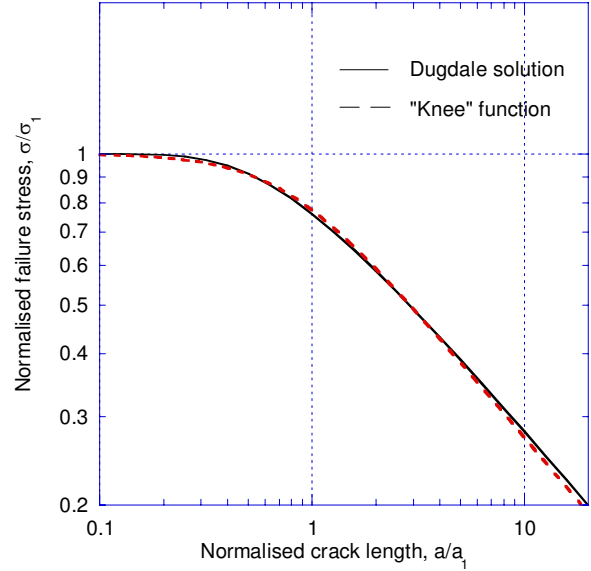} }
\caption{The comparison between the two-criterion strength scaling according to the Dugdale model (solid line) and the 'knee' function approximation (dashed line).}
\label{fig:four}
\end{figure}

For our present purposes we quote the following result from the Dugdale model \cite{dug}:
\begin{equation}
\frac{\sigma}{\sigma_1}=\frac{2}{\pi} \arccos \left[ \exp \left( -\frac{\pi E^* \delta_c}{8\sigma_1 a} \right) \right].
\end{equation}
Here we denote by $\sigma_1$ the limiting (yield) stress, $E^*$ is the plane strain Young's modulus, and $\delta_c$ is the critical crack tip opening displacement (material property). Introducing a characteristic crack length
\begin{equation}
a_1=\frac{\pi E^*\delta_c}{8\sigma_1},
\end{equation}
the failure criterion can be re-written in a convenient form allowing direct comparison with other expressions considered here,
\begin{equation}
\frac{\sigma}{\sigma_1}=\frac{2}{\pi}\arccos \left[ \exp \left( -\frac{a_1}{a} \right) \right].
\label{eq:dug}
\end{equation}
We now wish to see how well the two-criterion failure behaviour can be described by the use of the merging function, or 'knee' function, introduced above to describe Irwin's results in Fig.2 and Fig.3. We use the form 
\begin{equation}
\frac{\sigma}{\sigma_1}=\frac{1}{(1+(a/a_2)^n)^{1/(2n)}},
\label{eq:dugfun}
\end{equation}
which is different from that of equation (\ref{eq:irwinfun}) only in that the reference crack length $a_1$ has been replaced with $a_2$. The comparison between the Dugdale solution and the 'knee' function form in shown in Fig.4. Excellent agreement is found between the two descriptions, equations (\ref{eq:dug}) and (\ref{eq:dugfun}), with the following values of the parameters
\begin{equation}
n=1.996\pm 0.052,\quad \quad a_2/a_1=0.75\pm 0.01.
\end{equation}
Therefore, equation (\ref{eq:dug}) is also well approximated by the following simpler expression 
\begin{equation}
\frac{\sigma}{\sigma_1}=\frac{1}{\left(1+(24\sigma_1 a/3\pi E^*\delta_c)^2\right)^{1/4}}.
\end{equation}
However, equation (\ref{eq:dugfun}) offers greater flexibility, e.g. in terms of describing the sharpness of the transition: in fact, we have already seen (Fig.3) that the best choice of parameter $n$ for the description of Irwin's data is $n=10$, not $n=2$ that corresponds most closely to the Dugdale solution. 
It is also worth noting that the characteristic crack lengths in the two descriptions, $a_1$ for the Dugdale model and $a_2$ for the 'knee' function, are close in magnitude and only differ from each other by a multiplier of the order unity. This is only to be expected from the introduction of the 'knee' function in the previous section, since the parameter $x_0$ in equation (\ref{eq:knee}) has a value close to the vlaue of $x$ at which the switch between two power law behaviours takes place. 

\begin{figure}
\centerline{ \includegraphics[height=8.cm]{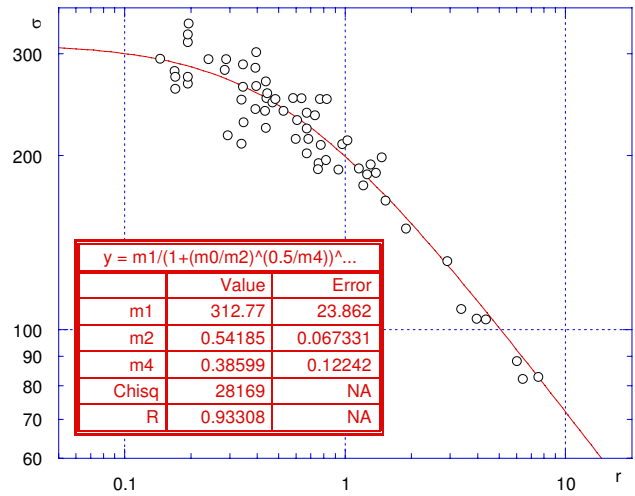} }
\caption{The experimental data of Mecholsky \cite{mech} on the scaling of fracture zones in a ceramic described using the 'knee' function.
}
\label{fig:five}
\end{figure}

Having succeeded in matching the Dugdale curve for ductile materials using the 'knee' function, we can turn our attention to the case of quasi-brittle materials. Mecholsky et al. \cite{mech} investigated the fracture surfaces of a ceramic (Pyroceram 9606) and identified a system of concentric annular zones emanating from the point of initiation, that were identified by the terms 'mirror', 'mist', and 'hackle', to indicate the varying degree of surface roughness. The variation in roughness is clearly associated with the change in crack propagation speed, which in turn is likely to depend on the amount of excess strain energy available for crack propagation and acceleration. Since this excess strain energy is clearly itself a function of crack size and applied stress, it is natural to find that the radii of zones exhibiting certain roughness scaled with the fracture stress. Mecholsky used a simple power law relationship,
\begin{equation}
\sigma r^m = M,
\end{equation}
where $r$ may refer to the radius of the 'mirror', 'mist' or 'hackle' region. For large zone sizes the above equation shows LEFM scaling with $m=0.5$. However, plotting the failure stress $\sigma$ versus the radius $r$ in bilogarithmic scale (Fig.5) reveals a clear deviation from 'pure' power law scaling at small radii. The set of experimental data in question can once again be successfully described using the 'knee' function (Fig.5).    

\section{\label{sec:k-t}Fatigue growth threshold and the Kitagawa-Takahashi diagram}

The transition between two power laws considered in the previous section in the context of static fracture strength is also observed, in virtually identical form, in the context of cyclic loading. Fatigue crack growth threshold is a term that refers to the stress level below which no crack growth is observed. It is found that, for nominally defect-free samples this threshold is expressed in terms of the stress range, $\Delta\sigma_0$, while for samples containing cracks the threshold value of the stress intensity factor range, $\Delta K_0$. Kitagawa and Takahashi \cite{kit} proposed combining the two criteria on a single diagram, thus producing a map on which non-propagating cracks occupy the region below both curves. The transition crack length $a_0$ is defined by the relation
\begin{equation} 
\Delta\sigma_0 \sqrt{\pi a_0}=\Delta K_0, \quad {\rm i.e.} \quad
a_0=\left( \frac{\Delta K_0}{\Delta\sigma_0} \right)^2.
\label{eq:kit}
\end{equation}

El Haddad \cite{elhad} noted that in the intermediate region the experimental data in fact lie some way below both curves, and suggested a form of the threshold stress range combining the two criteria given by
\begin{equation}
\Delta\sigma=\frac{\Delta K_0}{\sqrt{\pi(a+a_0)}},
\end{equation}
We note that this expression represents a particular case of the 'knee' function introduced here. Indeed, using the relationship of equation (\ref{eq:kit}) the above formula can be rewritten as
\begin{equation}
\Delta\sigma=\frac{\Delta\sigma_0}{(1+(a/a_0))^{1/2}}.
\label{eq:elhad}
\end{equation}
To illustrate the above discussion we use the fatigue threshold data compiled by Murakami and Endo \cite{murak}, Fig.6 and Fig.7. Instead of the El Haddad formula (\ref{eq:elhad}), a general form (\ref{eq:dugfun}) can be used with parameter $n$ optimised to provide the best match to the data. It is then found that the data is better described by the formula
\begin{equation}
\Delta\sigma=\frac{\Delta\sigma_0}{(1+(a/a_0)^{1/2})}=\frac{\Delta\sigma_0}{(1+\sqrt{a/a_0})}.
\end{equation}

\begin{figure}
\centerline{ \includegraphics[height=8.cm]{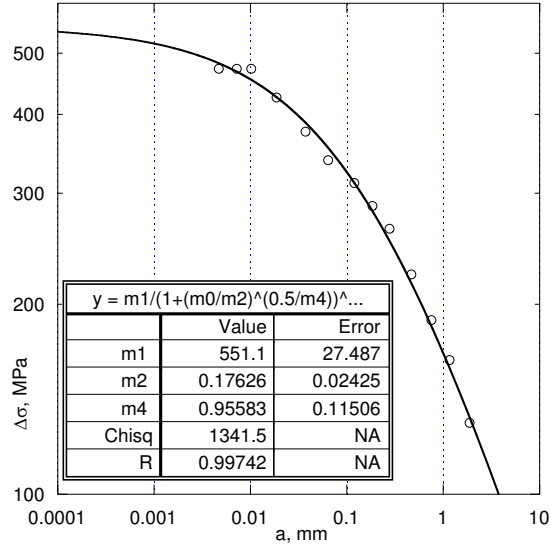} }
\caption{
The 'knee' function description of the fatigue threshold data compilation of Murakami and Endo \cite{murak}.
}
\label{fig:six}
\end{figure}

\begin{figure}
\centerline{ \includegraphics[height=8.cm]{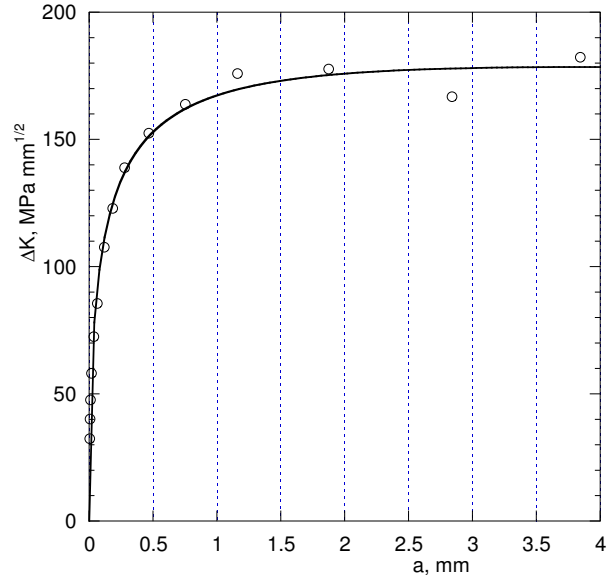} }
\caption{
The fatigue threshold data of Fig.6 plotted as stress intensity factor range $\Delta K$ versus defect size $a$.
}
\label{fig:seven}
\end{figure}

\section{\label{sec:paris}Kinetic diagrams of fatigue crack growth} 

Fig.8 shows the dependence of fatigue crack growth rate on applied stress intensity factor range, $\Delta K$, for a titanium alloy \cite{bot}. The data has been presented on bilogarithmic scale with the crack growth rate, $da/dN$, chosen as the abscissa. This choice allows the 'knee' function to be applied for the description of the dependence, sought in the form
\begin{equation}
\Delta K - \Delta K_{th}=\frac{\Delta K_{Ic}-\Delta K_{th}}{(1+(a'_0/a')^{bc})^{1/c}}.
\end{equation}
where $\Delta K_{th}$ denotes the fatigue threshold, $\Delta K_{Ic}$ corresponds to the critical stress intensity factor for fast fracture, $a'=da/dN$ is the crack advance per cycle, and $a'_0$ is the reference value of this parameter. Parameters $b$ and $c$ describe the slope of the diagram and the sharpness of the transition. Assuming stress intensity factor threshold of $\Delta K_{th}=7.5 {\rm MPa\sqrt{m}}$, the following values of the parameters are found
\begin{equation}
\Delta K_{Ic}=86\pm 7{\rm MPa\sqrt{m}}, \,
a_0=(6.1\pm 1.3)\times 10^{-6} {\rm m},
\end{equation}
corresponding to the continuous curve shown in Fig.8.

\begin{figure}
\centerline{ \includegraphics[height=8.cm]{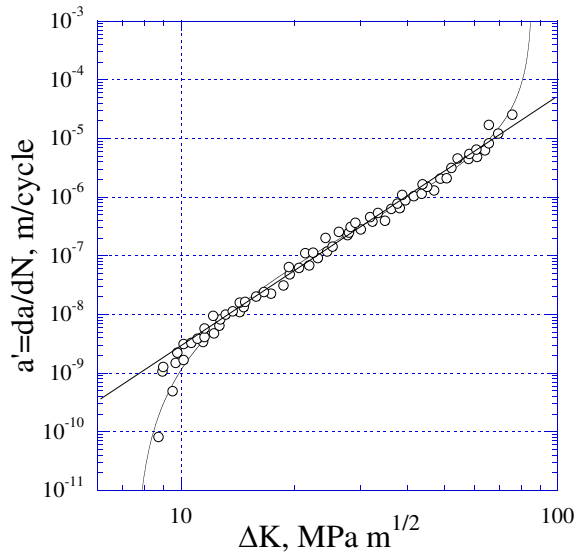} }
\caption{
Paris fatigue crack growth diagram from \cite{bot} described by the 'knee' function. The power law indicated by the straight line represents an intermediate asymptotic.
}
\label{fig:eight}
\end{figure}

In the central region between the asymptotics of large or small values of the stress intensity factor range the Paris diagram appears to contain a region of pure power law scaling. This phenomenon is referred to as an {\em intermediate asymptotic} \cite{baren}, and is illustrated by the straight line fit to the central region of the diagram. It is this relationship that became known as the Paris law. It is interesting to note that it persists over several decades of fatigue crack growth rates.

\begin{figure}
\centerline{ \includegraphics[height=8.cm]{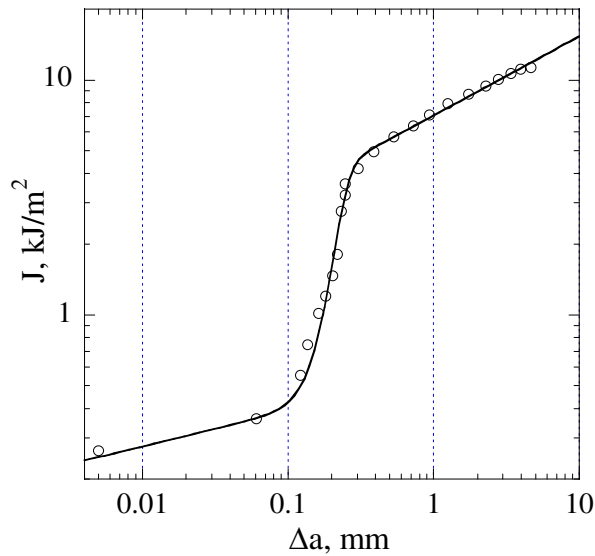} }
\caption{
A typical $J-R$ diagram (data from Miserez et al., \cite{miser})
}
\label{fig:nine}
\end{figure}

Crack resistance curves, or $R$-curves are often employed for the description of varying apparent strain energy release rate required for the onset and continuation of crack propagation. Consideration of such curves is interesting in the present context, since they capture the interaction between external loading and the multi-scale material response, beginning from microscopic tip blunting processes occurring at the immediate vicinity of the crack tip, but progressively involving larger regions of material in plastic deformation (and hence energy dissipation) as the load becomes higher, or the number of fatigue cycles is increased. The morphology and size of zones of plastic deformation and distributed damage play a central role in determining the scaling of observed strength. For the purposes of the present discussion, however, we concentrate on the discussion of the shape of crack resistance curves.

Miserez et al. \cite{miser} presented a careful study of crack propagation $R$-curves in terms of the $J$-integral ($J-R$ curves) for several high volume fraction particulate composites (aluminium matrix containing 50\% of $\rm Al_2O_3$ or $\rm B_4C$ particles). The authors identify three distinct stages of the process:
\begin{itemize}
\item{Initial part conventionally attributed to crack-tip blunting in unreinforced metallic alloys. Some apparent crack extension occurs in the composites in this study in this domain due to internal damage build-up at the crack tip.}
\item{A second stage starting close to the maximum load. This region is thought to correspond to macroscopic crack propagation. For finer particle sizes (10 or 5 $\mu$m) the crack often propagates in an unstable manner.}
\item{Fully stable crack propagation, under which a plateau value is finally attained in a region well beyond the domain of validity of $J$-controlled fracture.}
\end{itemize}

The above identification is based on mechanistic arguments. Fig.9 shows the data of Miserez et al. \cite{miser} re-plotted in bi-logarithmic scale and approximated by continuous 'knee' function curve of the type introduced here. This interpretation heps reveal a more detailed picture. In particular, the three distinct regions of power law scaling can be readily identified. Also apparent is the lack of a plateau in the extreme of large crack extensions $a$, and the significance of the range of lengths between 100 and 300 $\mu$m in the transition between different modes of material response. In the case considered this characteristic range of length scales corresponds to the transition from the plastic zone containing a small number of particles, each surrounded with locally yielded material, to the regime of extended flow with large numbers of particles absorbed in the plastic zone, and the material behaviour corresponding to that of structure-less continuum. 

\begin{figure}
\centerline{ \includegraphics[height=8.cm]{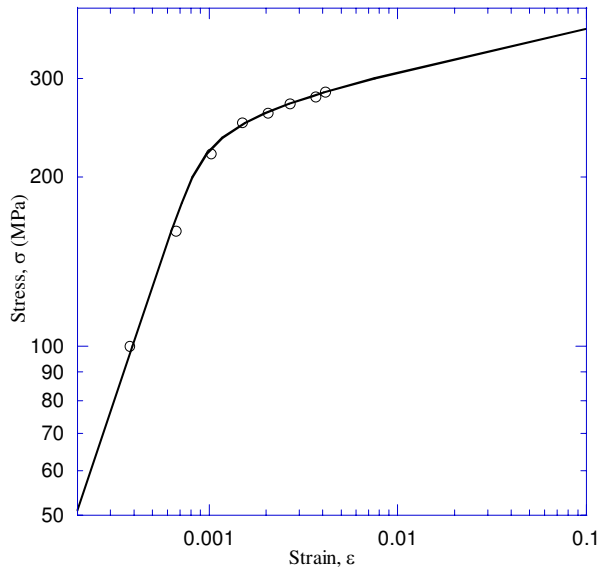} }
\caption{
The stress-strain curve of an elastic-plastic material obeying the Ramberg-Osgood law.
}
\label{fig:ten}
\end{figure}

\section{\label{sec:elplas}Stiffness of an elasto-plastic material}

In the analysis of constitutive laws of material deformation multi-scale power law behaviour plays a special role. In continuum deformation modelling the strain is assumed to be additive, i.e. the total strain is given by the sum of elastic and inelastic parts. In the case of time independent elasto-plasticity the Ramberg-Osgood law \cite{ram} is often used:
\begin{equation}
\varepsilon=\frac{\sigma}{E}+\left(\frac{\sigma}{K}\right)^m
\label{eq:ram}
\end{equation}
At small applied stresses the strain response is dominated by linear elasticity. However, when stress approaches a critical value (yield strain), the nature of the response changes, and for large stresses the strain is dominated by power law hardening. Ramberg-Osgood law for the description of this transition is popular both for its simplicity and due to the fact that it provides adequate approximation for a wide range of strains. We wish to explore the general background to this situation, and attempt to identify the functional form suitable for the description of the transition between power law asymptotics in elasto-plastic deformation.

For illustration we choose the Ramberg-Osgood description of stainless steel 316, for which the parameters were found to be $E=250\pm 21 {\rm GPa}, k=436\pm 15{\rm MPa}, m=13.4\pm 1.1$. The stress-strain curve for this material is shown in Fig.10 in bi-logarithmic coordinates, including the markers for experimental points and the continuous curve representing the relationship in equation (\ref{eq:ram}). For strains below about 0.001 linear elasticity is the dominant mechanism, while for strains greater than about 0.001 power law plasticity takes over as the principal mechanism determining total strain.

\begin{figure}
\centerline{ \includegraphics[height=8.cm]{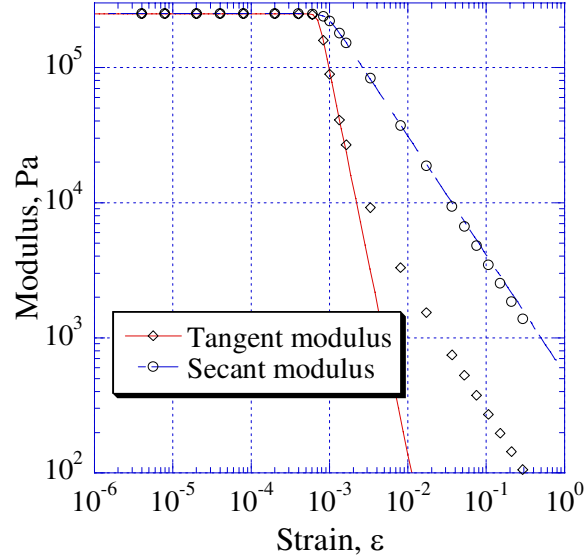} }
\caption{
The tangent and secant stiffness of an elastic-plastic material.
}
\label{fig:eleven}
\end{figure}

\begin{figure}
\centerline{ \includegraphics[height=8.cm]{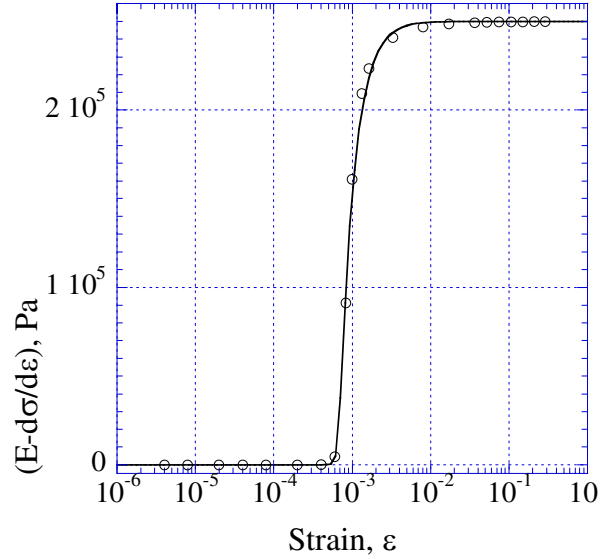} }
\caption{
The reduction of the instantaneous stiffness of an elastic-plastic material with respect to its Young's modulus as a function of strain.
}
\label{fig:twelve}
\end{figure}

The critical value of stress  $\sigma^*$ at which the elasto-plastic transition takes place is associated with yield within the framework of this model, and can be readily assessed approximately. Equating elastic part of the total strain to the plastic part the yield stress and strain values are found to be
\begin{equation}
\sigma^*=\left(\frac{k^m}{E}\right)^{1/(m-1)}, \quad
\epsilon^*=2\left( \frac{k}{E} \right)^{m/(m-1)}.
\end{equation}

The {\em tangent modulus}, or the instantaneous elasto-plastic stiffness of the material, is defined as $E^T=d\sigma/d\epsilon$. 
 
The {\em secant modulus} is defined as $E^S=\sigma/\epsilon$, and can be expressed as
\begin{equation}
\frac{E^S}{E}=\frac{1}{1+(\sigma/\sigma^*)^m},
\end{equation}
that is, is a particular form of the 'knee' function. It is possible to use the general form of the knee function (\ref{eq:knee}) to describe the dependence of the secant modulus on the total strain. The result is shown in Fig.11, together with the curve for the tangent modulus. 

In order to study the variation of tangent modulus in more detail we introduce a compound parameter describing the magnitude of deviation of the tangent modulus from its value in the elastic regime, $E'=E-E^T$. The dependence of modulus deviation, $E'$, on strain is shown in Fig.12, together with the continuous curve illustrating the quality of the 'knee' function fit to the data. Of particular interest here is the sharpness of transition in the vicinity of yield, and the suitability of the 'knee' function for the description of this phenomenon. 

\begin{figure}
\centerline{ \includegraphics[height=8.cm]{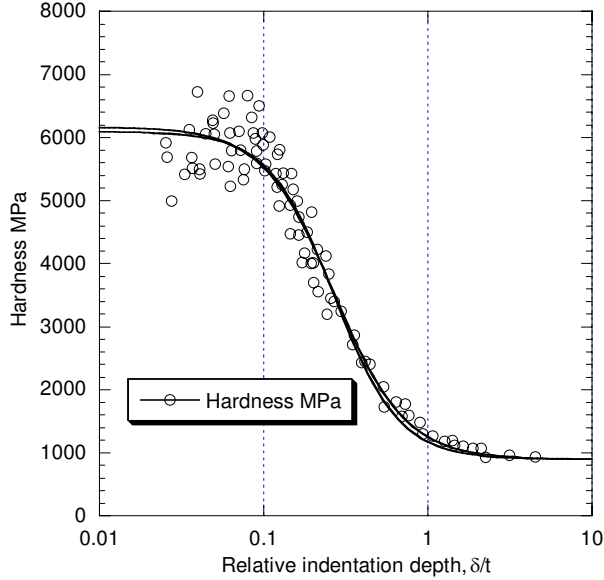} }
\caption{
The variation of apparent hardness of a coated system as a function of the relative indentation depth.
}
\label{fig:thirteen}
\end{figure}

\section{\label{sec:hardness}Hardness of coated systems}

A particular example of indentation size effect on hardness that we wish to consider here concerns the variation of the apparent hardness of a coated system with the indentation depth. A model for the description of this variation was introduced by Korsunsky et al. \cite{woi} in the form 
\begin{equation}
H-H_s=\frac{H_f-H_s}{1+(\delta/\delta_0)^m},
\end{equation}
where $H$ is the apparent hardness, $H_f$ is the intrinsic hardness of the coating film, $H_s$ is the substrate hardness, $\delta=d/t$ represents the relative indentation depth, where $d$ is indentation depth and $t$ is film thickness; and $\delta_0$ is the reference value of the indentation depth. The above formula represents a particular case of the 'knee' function.  

\begin{figure}
\centerline{ \includegraphics[height=8.cm]{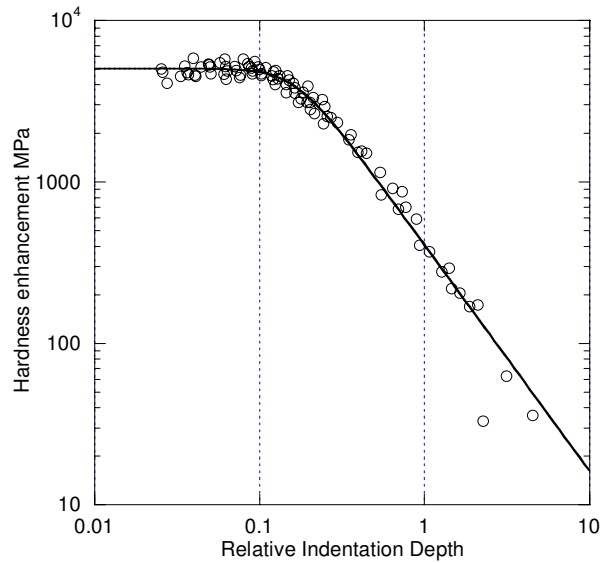} }
\caption{
The 'knee' function description of the hardness increase due to the presence of the coating.
}
\label{fig:fourteen}
\end{figure}

The application of the 'knee' function description is illustrated in Fig.13 and Fig.14 using the vast body of indentation data collected from electroplated Ni coatings on Cu by Tuck et al \cite{tuck}. Fig.13 shows the apparent system hardness data in semi-logarithmic coordinates, while Fig.14 reveals the multi-scaling power law transition by plotting the hardness increase over the substrate, $H-H_s$, against the relative indentation depth $\delta$, in a bi-logarithmic plot.

\section{\label{sec:creep}Discussion and conlusions}

Data and analysis presented in this paper cover a broad range of experimental observations from various branches of the science of material strength. The coverage is not intended to be complete, as this would never be possible in a short paper. Phenomena not dwelt upon here, but known to obey similar laws, include time-dependent plasticity (creep), grain size dependence of yield stress (Hall-Petch law), etc. For a more complete analysis the reader is referred to the relevant sections of a fuller report \cite{korbot}.

It is also important to note that more complex situations than those considered here may be encountered, such as discontinuous transitions between power law regimes, dependence on multiple parameters, etc. It is hoped, however, that the ideas presented here may be found helpful in developing better interpretative tools for other, more complex situations.

\section*{Acknowledgements}

The author would like to thank Professor L.R. Botvina for her most valuable critical comments on the ideas presented in this paper. 



\end{document}